\documentstyle [12pt]{article}
\begin{document}
\centerline {Fourth International Workshop on Theoretical and
Phenomenological Aspects}
\centerline {~~~~~~~~~~~of Underground Physics, Toledo (Spain)
September 17-21 1995}
\vskip 1cm
\centerline {\bf COSMOLOGICAL IMPLICATIONS OF A POSSIBLE CLASS OF}
\vskip 3mm
\centerline {\bf PARTICLES ABLE TO TRAVEL FASTER THAN LIGHT
(abridged version)}
\vskip 1.5cm
\centerline {\bf L. GONZALEZ-MESTRES}
\vskip 4mm
\centerline {Laboratoire d'Annecy-le-Vieux de Physique des
Particules,}
\centerline {B.P. 110 , 74941 Annecy-le-Vieux Cedex, France}
\vskip 1mm
\centerline {and}
\vskip 1mm
\centerline {Laboratoire de Physique Corpusculaire,
Coll\`ege de France,}
\centerline {11 pl. Marcellin-Berthelot, 75231 Paris Cedex 05 ,
France}
\vskip 1.5cm
Special relativity
is usually presented
as an intrinsic property of space and time: any material body moves
inside a unique minkowskian structure governed by Lorentz
transformations.
Geometry is also the startpoint of
the theory of gravitation in general relativity and provides its
ultimate dynamical concept.
However, a look to various dynamical systems would suggest a more
flexible view with
the properties of matter playing the main role.
In a two-dimensional galilean space-time,
a sine-Gordon soliton of the form:
\equation
\phi _v (x,t) =
4~arc~tan~[exp~(\pm~\omega c_o^{-1}~(x-vt)~(1- v^2/c_o^2)^{-1/2})]
\endequation
moving at speed $v$ has size
$\Delta x$ $=$ $c_o\omega ^{-1}~(1- v^2/c_o^2)^{1/2}$ ,
proper time $d\tau $ $=$ $dt~(1- v^2/c_o^2)^{1/2}$ ,
energy $E$ $=$ $E_o~(1- v^2/c_o^2)^{-1/2}$ with $E_o$ $=$
rest energy, mass $m$ $=$ $E_o/c_o^2$
and momentum $p$ $=$ $mv~(1- v^2/c_o^2)^{-1/2}$ .
The constant $c_o$ depends on the underlying dynamics and plays
the role of a critical speed, just as $c$
(the speed of light) is the critical
speed in vacuum.
Everything looks perfectly "minkowskian" and a "sub-world"
restricted to such solitons would feel a "relativistic"
space-time, with $c_o$ instead of $c$ . In the real vacuum,
if Lorentz invariance is just a property of equations
describing a sector of matter at a given scale, an absolute
rest frame may exist and $c$ will not necessarily be the only critical
speed [1] . Similar situations have already been found in physics:
in a perfectly transparent crystal it is possible
to identify at least two critical speeds (those of sound and light).
Superluminal sectors may exist in
vacuum related to new degrees of freedom not yet unraveled
experimentally.
\vskip 3mm
The new particles would not be tachyons: they may feel different
minkowskian space-times with critical speeds $c_i$ $\gg $ $c$
(the subscript $i$ stands for the $i$-th sector) and
behave kinematically like ordinary particles. The "ordinary"
sector will contain "ordinary" particles with critical speed $c$ .
Each sector will have its own Lorentz invariance,
but interactions between two different sectors
will break both Lorentz invariances.
Superluminal particles will have [1] rest energy
$E_{rest} = mc_i^2$ , for inertial mass $m$ and critical speed $c_i$~.
Energy and momentum conservation will not be spoiled by the
existence of several critical speeds in vacuum:
conservation laws will
as usual hold for phenomena leaving the vacuum unchanged.
If superluminal particles couple weakly to ordinary matter,
their effect on the ordinary sector may occur basically
at high energy and short distance, far from conventional
tests of Lorentz invariance.
As the graviton is an "ordinary" gauge particle associated to the
local Lorentz invariance of the ordinary sector,
superluminal particles will not necessarily couple to gravity
in the usual way. Concepts so far considered as
very fundamental (i.e. the universality of the exact equivalence
between inertial and gravitational mass)
will fail and become approximate sectorial properties.
Gravitational properties of vacuum remain unknown and new forces
may have governed the expansion of the Universe.
\vskip 3mm
If superluminal sectors exist and Lorentz invariance is only a
sectorial property, the Big Bang scenario may become quite different.
For a sector with critical speed $c_i$ and apparent Lorentz
invariance at distance
scales larger than $k_i^{-1}$ , where $k_i$ is a critical wave
vector scale, we expect
the appearance of a critical temperature $T_i$
given approximately by:
\equation
k~T_i~\approx ~\hbar ~c_i~k_i
\endequation
where $k$ is the Boltzmann constant and $\hbar $ the Planck constant.
Above $T_i$ , the vacuum will not necessarily allow for particles
of the $i$-th sector as excitations. If $k_o$ stands for the
critical wave vector scale of the ordinary sector,
above $T_o$ $\approx $ $k^{-1}\hbar ck_o$
the Universe may
have contained only superluminal particles
whereas superluminal and ordinary
particles coexist below $T_o$ . It may also happen that some ordinary
particles exist above $T_o$ , but with different
properties (like sound above the melting point).
Ordinary particles did most likely not govern the beginning of
the Big Bang
and dynamical correlations appear to have been able to propagate
must faster than light
in the very early Universe.
At lower temperatures, the effect remains but is limited by
annihilation, decoupling and "Cherenkov radiation"
(emission of ordinary particles by
the superluminal ones at $v$ $>$ $c$) in vacuum.
\vskip 3mm
The existence of superluminal particles seems potentially able
to invalidate standard arguments leading to the
so-called "horizon problem" and "monopole problem", because:
a) above $T_o$ , correlations propagate mainly at superluminal speed;
b) below $T_o$ , the annihilation of superluminal
particles into ordinary ones is expected to release very large
amounts of kinetic energy from the rest masses and generate
a fast expansion of the Universe. If $kT_o$ is not higher
than $\approx $ 10$^{14}$ $GeV$
($k_o$ $\approx $ 10$^{-27}$ $cm$ , time scale
$\approx $ 10$^{-38}$ $s$), the formation of the symmetry-breaking
condensate in vacuum may even have occurred above $T_o$ . Because of
superluminal degrees of freedom and of the expected phase
transitions at $T_o$ and at all $T_i$ ,
it seems difficult to set any "natural time scale"
(e.g. at the Planck time $t_p$ $\approx $ 10$^{-44}$ $s$ ).
Arguments leading to the "flatness" or "naturalness"
problem, as well as
the role of gravity and the concept of the cosmological constant,
should
be reconsidered.
\vskip 5mm
{\bf Reference}
\vskip 4mm
\noindent
[1] L. Gonzalez-Mestres, "Properties of a possible class of particles
able to travel faster than light", Proceedings of the
Moriond Workshop on "Dark Matter in Cosmology, Clocks and Tests of
Fundamental Laws", Villars (Switzerland), January 21 - 28 1995 ,
Ed. Fronti\`eres. Paper astro-ph/9505117 of electronic library.
\end{document}